# Spacing Dependent and Doping Independent Superconductivity in Intercalated 1T Two Dimensional SnSe$_2$

Hanlin Wu[1*], Sheng Li[1*], Michael Susner[2], Sunah Kwon[3], Moon Kim[3], Timothy Haugan[2], Bing Lv[1†]

[1] *Department of Physics, The University of Texas at Dallas, Richardson, TX 75080, USA*

[2] *Air Force Research Laboratory, Aerospace Systems Directorate, Wright-Patterson AFB, OH 87123, USA*

[3] *Department of Materials Science and Engineering, University of Texas at Dallas, Richardson, TX 75080, USA*

The weak van der Waals interlayer interactions in the transition metal dichalcogenide (TMD) materials have created a rich platform to study their exotic electronic properties through chemical doping or physical gating techniques. We reported bulk superconductivity up to 7.6 K through careful manipulation of the charge carrier density and interlayer spacing $d$ in the chemically intercalated two dimensional 1T-SnSe$_2$ phase. We found, for the first time in the two dimensional SnSe$_2$, that polar organic molecules cointercalated with the alkali metal Li into the basal layers, thus significantly enhancing the superconducting $T_c$. We observed that the $T_c$ scales with the basal spacing distance, meanwhile being almost independent of $x$ in Li$_x$(THF)$_y$SnSe$_2$ system. Our results offers a new general chemical route to study the rich electron correlations and the interplay of charge density wave and unconventional superconductivity in the two dimensional material.

## I. INTRODUCTION

Two dimensional layered materials such as transition metal dichalcogenides (TMD) have attracted significant research interests in the past decade due to their fantastic physical properties such as layer-dependent band gap [1-3], topology [4-9], and valley-related transport induced by broken inversion symmetry [10-14]. Great progress has been made in the fundamental and nanoscale device studies of these materials towards their potential applications in electronics, optoelectronics, and spintronics [15-18]. The layered feature of these TMDs permits a variety of guest atoms, molecules or electrons to be intercalated into the van der Waals gap of the host materials. Numerous works have reported on the drastic change to the optical and electrical properties of layered materials when they are subjected to intercalation into the van der Waals gaps with group I or II alkali metal atoms or inorganic molecules [19-25]. Superconductivity has been induced in many of these TMD materials through both chemical doping/intercalation [26-31] and physical gating methods [32-36], with the very recent highlight of gate-induced superconductivity in the topological WTe$_2$ system [37,38].

Comparatively, much fewer efforts have been made in the study of post-transition metal dichalcogenides, which have different electronic structures from TMDs owing to their lack of $d$-electron contributions to bands near the Fermi surface [39,40]. Tin-based dichalcogenides, for example, crystallized solely in the two dimensional CdI$_2$-type 1T phase to date [41], similar as MoS$_2$, where hexagonal closely packed sandwich layers are stacked with a periodicity of one layer tin atom and two layers chalcogenide atoms forming tilted SnSe$_6$ octahedra. The interlayer spaces of ~ 5.785 Å for SnS$_2$ and 6.137 Å for SnSe$_2$ that are loosely bonded by van der Waals forces. Compared to MoS$_2$, the larger interlayer spacing in the Sn dichalchogenide materials will result in a higher possible degree of chemical species intercalations as well as provide a buffer for the large volume change associated with the intercalation processes. In fact, superconductivity has been reported through the intercalation of organic metallocenes with possible charge density wave (CDW) interplay at higher temperatures, another signature for unconventional superconductivity [42-44]. Gating techniques have induced superconductivity at 3.9 K with the in plane upper critical field greatly exceed the Pauli limit [45]. In addition, interface superconductivity is suggested in a SnSe$_2$/graphite heterostructure through STM study [46,47]. The unexpected large superconducting gap bears a number of similarities to that of cuprates, providing additional strong evidence for unconventional superconductivity. In addition, SnSe$_2$ and SnS$_2$, are also widely used for solar cells [48], anode of lithium/sodium ion batteries [49,50], field effect transistors [51] and also photodetector [52].

Herein, we report bulk superconductivity up to 7.6 K in the alkali metal Li-intercalated Li$_x$SnSe$_2$ and the organic molecule cointercalated Li$_x$M$_y$SnSe$_2$, where M is tetrahydrofuran (THF) or propylene carbonate (PC). We have systematically investigated the controllability of both carrier density $x$, organic molecules M, and the interlayer spacing $d$. While an interesting $x$-independent superconducting transition temperature is preserved, a significant $T_c$ enhancement of up to 100%, depending on the interlayered spacing, was observed by organic molecule cointercalation. Finally, detailed transport studies

* These authors contributed too this work
† To whom correspondence should be addressed: blv@utdallas.edu

suggest the presence of unconventional superconductivity in this cointercalated SnSe$_2$ system.

## II. EXPERIMENTAL

The SnSe$_2$ crystals was synthesized as follows. the mixture of high purity tin ingot (Alfa Aesar, 99.99%) and selenium pieces (Alfa Aesar, 99.999%),were sealed in a evacuate quartz tube heated to 950 °C in 20 hours for 24 hours, and then slowly cooled down to 650 °C to obtain high purity SnSe$_2$ precursor. The high quality SnSe$_2$ single crystals are grown through vertical Bridgeman method using the pre-reacted SnSe$_2$ powder in the doubled sealed quartz tube container, with growth zone at 950 °C, cooling zone at 500 °C, moving rate at 1.5 mm/h and total growth time of 20 days. The grounded powder of SnSe$_2$ crystals grown from Bridgeman method was used as starting material for our intercalation studies.

Since the Li-intercalated samples are air sensitive, all the intercalation and co-intercalation process were performed inside purified Ar-atmosphere glovebox with total O$_2$ and H$_2$O levels < 0.1 ppm. The $n$-butyllithium ($n$-BuLi) used was 2.6 M (1 M = 1 mol/L) concentration in hexane solution from Alfa Aesar and diluted to 0.02 - 0.2 M. Lithium naphthalene solutions (Li-Naph) with concentration of 0.02 - 0.2 M were prepared by dissolving equimolar amounts of naphthalene and Li metal into THF while magnetic stirring and heating at ~ 70 °C. Organic solvents hexane, THF and PC are vacuum distilled and dried with molecular sieve before usage. We used several different intercalation methods to prepare a variety of samples.

(i) Li$_x$SnSe$_2$ samples were prepared by soaking the parent SnSe$_2$ powder in various diluted 0.02 - 0.2 M n-BuLi solution. The n-BuLi contained in the solutions is more than three equivalent of actual doping level of SnSe$_2$ to avoid the dilution of the n-BuLi in the intercalation process. (*cf.* Figure 1a).

(ii) The cointercalated Li$_x$(THF)$_y$SnSe$_2$ were prepared by soaking in various concentration (0.02 - 0.4 M) Li-Naph solution in THF with a typical period for 1 day under magnetic stirring. (*cf.* Figure 1c).

(iii) Since the Li metal does not reacts with naphthalene nor dissolves in the PC solvent, the direct PC cointercation using method (ii) is not feasible. Alternatively, the PC cointercalated Li$_x$(PC)$_y$SnSe$_2$ were prepared by soaking Li-intercalated sample (by method i) in enough PC solution under magnetic stirring. (*cf.* Figure 1c).

The chemical composition, i.e Li concentration ($x$) in the products was determined within an accuracy of ±0.01 by inductively coupled plasma mass spectroscopy (ICP-MS) using Agilent Technologies 7900. Thus determined $x$ values are used throughout the manuscript.

The powder samples were dried, and cold pressed into pellets under uniaxial stress for the following physical measurements. X-ray diffraction patterns were collected using the Rigaku Smartlab. Electrical resistivity ρ was measured by employing a standard 4-probe method down to 1.8 K in a Quantum Design Physical Property Measurement System (PPMS). The magnetization was carried out using the Quantum Design Magnetic Property Measurement System (MPMS).

## III. RESULTS AND DISCUSSION

Because of its two dimensional layered nature, SnSe$_2$ thin flakes obtained from grinding the Bridgman-grown crystals have very strong c-axis preferred orientation in their X-ray diffraction (XRD) patterns. The strongest (001) peak, therefore, is used to compare and determine the lattice expansion during the intercalation process. In Figure 1 we shown the XRD patterns for both the intercalated samples and the parent compound. The use of different organic solvents can severely impact the intercalation process. The polar organic molecules such as THF and PC, with their stronger reducing power due to the cation-dipole interaction, can cointercalate with the cations, but non-polar solvent molecules such as hexane used in *n*-BuLi reaction are not cointercalated.

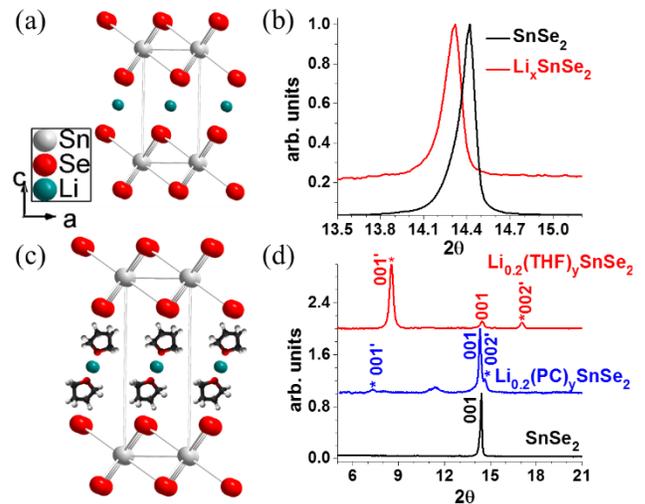

FIG. 1. (a) Schematic structural model for Li-intercalated Li$_x$SnSe$_2$; (b) (001) reflection peak of XRD pattern for selected SnSe$_2$ and Li$_{0.2}$SnSe$_2$ samples; (c) Schematic structural model for organic cointercalated Li$_x$(THF)$_y$SnSe$_2$; (d) (001) reflection peak of XRD pattern for SnSe$_2$, Li$_{0.2}$(THF)$_y$SnSe$_2$, and Li$_{0.2}$(PC)$_y$SnSe$_2$ samples. The Li content in the Li$_{0.2}$SnSe$_2$, Li$_{0.2}$(THF)$_y$SnSe$_2$, and Li$_{0.2}$(PC)$_y$SnSe$_2$ samples was determined through inductively coupled plasma mass spectroscopy (ICP-MS).

As seen in Figure 1a-b, in the *n*-BuLi/hexane process from method (i), Li atoms are solely intercalated into the interlayer space. Due to the small size and relatively small doping levels of Li ions, we observe only a small lattice shift (~ 0.1°) with only Li intercalation. Such shift indicates $c$ lattice parameters expands very slightly from 6.137 Å in parent SnSe$_2$ to 6.180 Å in the intercalated Li$_{0.2}$SnSe$_2$.

On the other hand, as seen in Figure 1c-d, the lattice parameter $c$ expands dramatically with the presence of cointercalated organic THF and PC molecules synthesized following processes (ii) and (iii). The basal spacing of the compound has been increased from 6.137 Å from SnSe$_2$, to 10.341 Å in the Li$_{0.2}$(THF)$_y$SnSe$_2$ compound. There is some small amount (< 10%) of SnSe$_2$ that remains unreacted during the intercalation which does not reduce with longer intercalation times nor slightly elevated reaction temperatures (e.g. using a hot plate). For the Li$_x$(PC)$_y$SnSe$_2$ sample, the basal spacing is further increased to 12.081 Å due to the stronger cation-dipole interaction of PC comparing to THF (dipole moment is 4.94 D for PC, and 1.63 D for THF, 1D= 3.3356 x 10$^{-30}$ Cm) [53]. However, this cointercalation process is quite inefficient, with large amount of Li$_x$SnSe$_2$ from $n$-BuLi/hexane left, shown in Figure 1d. The peak at 11.4°, corresponding to a 7.79 Å basal spacing, suggests that a possible different intercalation stage, i.e phase separation, exists in the particular sample.

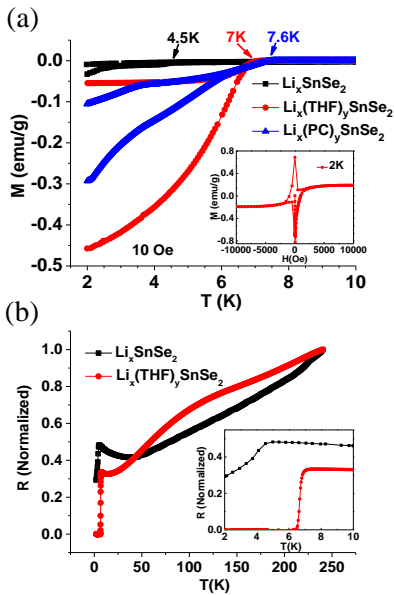

FIG. 2. (a) Magnetic susceptibility data for of Li and cointercalated SnSe$_2$ samples, where all three samples show diamagnetic signals; (b) Normalized temperature dependent resistivity data. The inset of (a) is the MH loop of Li$_{0.2}$(THF)$_y$SnSe$_2$ at 2 K, and inset of (b) is the enlarged resistivity data highlighting superconducting transition at low temperature from 2 - 10 K.

The cointercalation of organic THF and PC molecules is further evidenced by Thermogravimetric Analysis (TGA) measurements (see supplemental materials). Li$_{0.2}$(THF)$_y$SnSe$_2$ starts to decompose at ~ 50 °C through the loss of THF and Li$_{0.2}$(PC)$_y$SnSe$_2$ begins decomposition at 80°C; in both cases these reactions are endothermic in nature. However, given the presence of unreacted SnSe$_2$, Li$_x$SnSe$_2$, and the additional issue of possible phase separation, it is rather difficult to determine the precise amount of organic species in the cointercalated samples.

The $dc$ magnetization of intercalated Li$_{0.2}$SnSe$_2$, Li$_{0.2}$(THF)$_y$SnSe$_2$, and Li$_{0.2}$(PC)$_y$SnSe$_2$ are shown in the Figure 2a. We can see a clear enhancement of the transition temperature as the basal spacing is increased after intercalation/cointercalation processes. In the Li$_{0.2}$SnSe$_2$ sample, a small diamagnetic shift emerges from ~ 4 K under zero field cool (ZFC) with a very small shielding fraction, suggesting the presence of non-bulk superconductivity in the sample. In contrast, for Li$_{0.2}$(THF)$_y$SnSe$_2$, a much large diamagnetic signal from 7 K is observed. In addition, the M-H curve, shown in Figure 2a, suggests that type-II superconductivity is present in this sample. As previously noted, the Li$_{0.2}$(PC)$_y$SnSe$_2$ sample exhibits a further increased basal spacing. Further buttressing the relationship between basal spacing and $T_c$, we note a concomitant increase in the superconducting transition temperature in Li$_{0.2}$(PC)$_y$SnSe$_2$, with a further enhancement to 7.6 K. However, in the Li$_{0.2}$(PC)$_y$SnSe$_2$ sample, both zero field cooling (ZFC) and field cooling (FC) data show an additional field expulsion below 4 K, which could be caused by Li$_x$SnSe$_2$ or phase separation caused by another stage of intercalation.

The superconducting behavior in these samples can be further verified by the resistivity measurement shown in the Figure 2b. The resistivity measurements confirm the superconducting temperatures of 4.5 K and 7 K for Li$_{0.2}$SnSe$_2$ and Li$_{0.2}$(THF)$_y$SnSe$_2$, respectively, consistent with the magnetic measurements. For Li$_{0.2}$SnSe$_2$, only a 50% resistivity drop is observed, supporting the hypothesis of non-bulk superconductivity in the sample. For Li$_{0.2}$(THF)$_y$SnSe$_2$, the superconducting transition is quite sharp and the resistivity reaches to zero at 6.4 K with a transition width <0.6 K. The Li$_{0.2}$(PC)$_y$SnSe$_2$ sample is very difficult to compress as a small pellet after intercalation. Therefore, no resistivity measurement is attempted for this sample. It is worthwhile to mention that a clear hump in the $\rho(T)$ curve is present at ~ 100 K for Li$_{0.2}$(THF)$_y$SnSe$_2$, reminiscent to previous investigations of metallocene-intercalated SnSe$_2$ [44,54]. Similar as the other TMDs such as NbSe$_2$, this hump might associate with the charge density wave (CDW) transition, as a result of two dimensional Fermi surface nesting. Because the Li content, i.e. charge carrier changes, is the same for the Li$_{0.2}$SnSe$_2$ and Li$_{0.2}$(THF)$_y$SnSe$_2$ samples, this suggests that CDW, rather than charge carrier density, may play an important role for the occurrence of bulk superconductivity, and enhanced $T_c$ comparing with the gating technique.

Due to the presence of non-bulk superconductivity in Li$_x$SnSe$_2$ and the second transition noted in the Li$_x$(PC)$_y$SnSe$_2$ samples, we focused our studies of the carrier change effects on the Li$_x$(THF)$_y$SnSe$_2$ sample. The amount of lithium uptake by SnSe$_2$ could be controlled by changing the amount of Naph.-Li used. Indeed, the ICP-MS measurement has shown we obtained different x=0.01 – 0.30 samples through this method. On the other hand, no significant basal spacing increase is observed from the XRD patterns, indicating the amount of THF uptake is nearly the same for all these Li$_x$(THF)$_y$SnSe$_2$ samples with different Li content since the THF is abundant as the solvent. We show the temperature dependent resistivity data on these samples in Figure 3a. SnSe$_2$ is semiconducting with an indirect band gap 1.07 eV [40]. As the Li doping increases, the progressive suppression of the insulting behavior can be clearly seen. A slight resistivity drop is observed for $x$ = 0.03 even though the overall bulk material is still insulating, suggesting the emergence of superconductivity at this composition. Upon further Li doping, the magnitude of the superconducting diamagnetic response increases and the $\rho(T)$ data show true zero-resistivity behavior. However, both resistivity and magnetic measurements show that the $T_c$ is *more or less independent* of the Li doping level $x$ for $x > 0.03$ throughout of the region, in contrast to the conventional superconductivity. Taking the conventional cross point of the two slopes as the superconducting transition temperature $T_c$ from the resistivity and ZFC magnetic data, our summarized basal spacing, and doping level $x$ dependent $T_c$ is shown in the Figure 4.

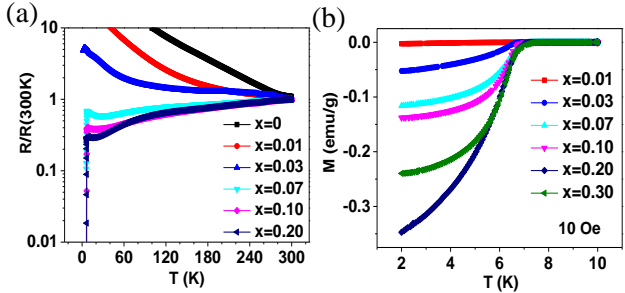

FIG. 3. (a) Normalized resistivity data of $Li_x(THF)_ySnSe_2$ with different Li concentrations from $0 \leq x \leq 0.2$; (b) Zero field cooled (ZFC) magnetization data for different $Li_x(THF)_ySnSe_2$ under 10 Oe field. The Li content is determined through the ICP-MS method.

We observed a monotonic increase in $T_c$ with basal spacing, in the Figure 4a. However, this trend does not completely rule out the possibility of a dome-like $T_c$ vs basal spacing behavior upon further intercalation, as seen in the intercalated ZrNCl system [55]. The second trend is that $T_c$ is independent of Li concentration for $x > 0.03$, shown in the Figure 4b. This behavior is very similar to the intercalated ZrNCl/HfNCl system where it was attributed to the $x$-independent $N(0)$ characteristics of a nearly-free electron two dimensional system [56-59]. Charge fluctuations associated with the CDW transiton might therefore play an important role to induce superconductivity in these intercalated systems.

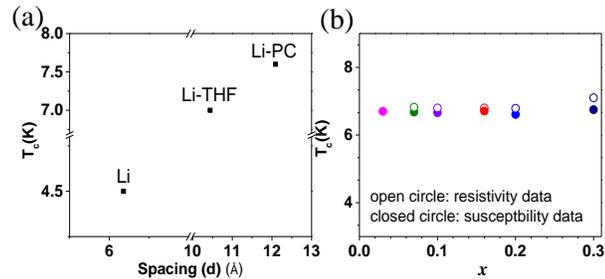

FIG.4. (a) Superconducting $T_c$ values with respect of the basal spacing distances; (b) $T_c$ values plotted as a function of doping level in $Li_x(THF)_ySnSe_2$ showing $x$ independent $T_c$.

In Figure 5 we show the field dependent resistivity data of $Li_{0.2}(THF)_ySnSe_2$. The noise near room temperature is likely associated with the condensation of moisture in the samples, since the sample is very air sensitive. The small upturn in $\rho(T)$ clearly visible prior to the superconducting transition can beascribed to the Anderson localization effect caused by inhomogeneity and disorder in the cointercalated samples [60].

Upon the application of the magnetic field, the superconducting transition is both suppressed and broadened, as expected. However, when we plot the upper critical field $H_{c2}$ as a function of temperature (Figure 5) we note a very steep increase in the value of $H_{c2}$ with temperature. Using a simple linear extrapolation, the zero-temperature upper critical field, $\mu_0H_{c2}(0)$, is estimated to be 9 T. In contrast, if we apply the Werhmer-Helfand-Hohenberge (WHH) model where only the orbital effect is taken into account, the estimated $\mu_0H_{c2}(0)$ is about 2.3 T. Such upper critical field behavior indicates increased anisotropy and likely multiband superconductivity nature in this system.

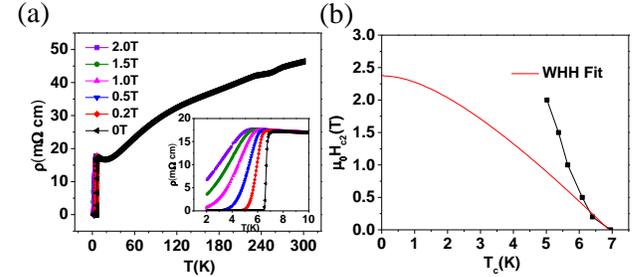

FIG. 5. (a) Field dependence of resistivity data and enlarged view of low temperature region in insert for $Li_{0.2}(THF)_ySnSe_2$ sample; (b) Upper critical field determined from resistivity data. The red curve is fitting curve from WHH model.

Specific heat data were performed to get further insight of the superconductivity. Unfortunately we did not observe any clear superconducting transition at 7 K for zero field data. By subtracting the data of zero field to that of high magnetic field, we can only see a very tiny specific jump. We compared the specific data of parent compound and the superconducting sample, shown in the Fig. 5. By Debye fitting of the data using $C = \gamma_N T + \beta T^3$, we can get $\gamma_N \sim 0$ and $\beta = 1.016$ mJ/mol K$^4$ and $\gamma_N = 0.654$ mJ/mol K$^2$ and $\beta = 1.228$ mJ/mol K$^4$, which corresponds to the electronic Sommerfeld coefficient and Debye temperature, for parent and superconducting samples respectively. The Debye temperature can be derived from the $\beta$ value through the relationship $\theta_D = (12\pi^4 k_B N_A Z/5\beta)^{1/3}$, where $N_A$ is the Avogadro constant, and Z is the number of atoms in the molecule. The calculated Debye temperature is 180 K and 166 K for parent and superconducting compounds respectively.

As the Sommerfeld coefficient corresponding to the density of states at Fermi level, with $\gamma_N \sim 0$ corresponding to the insulating properties of the parent compound. Once we introduce electrons through chemical intercalation, a small value of Sommerfeld coefficient $\gamma_N = 0.654$ mJ/mol K$^2$ will be introduced. If we take this value to $\Delta C/ \gamma_N T_c = 1.43$ for the BCS theory, the specific jump $\Delta C/ T$ at the transition temperature is 0.94 mJ/mol K$^2$, only 2% of the specific heat value at $T_c$, which could be a possible reason we did not observe the specific jump. Similar small Sommerfeld coefficient is also observed in the $Li_xZrNCl$ system. If we compare $\gamma_N$ and $T_c$ for $Li_xZrNCl$ and $Li_x(THF)_ySnSe_2$, with $\gamma_N \sim 1$ mJ/mol K$^2$ and $T_c \sim 12$ K for $Li_xZrNCl$ and $\gamma_N \sim 0.65$ mJ/mol K$^2$ and $T_c = 6.9$ K for $Li_x(THF)_ySnSe_2$, almost the same ratio between $\gamma_N$ and $T_c$ is obtained. The close relationship between $Li_xZrNCl$ and Li$_x$(THF)$_y$SnSe$_2$ may indicate unconventional superconductivity of the co-intercalated SnSe$_2$ [61,62].

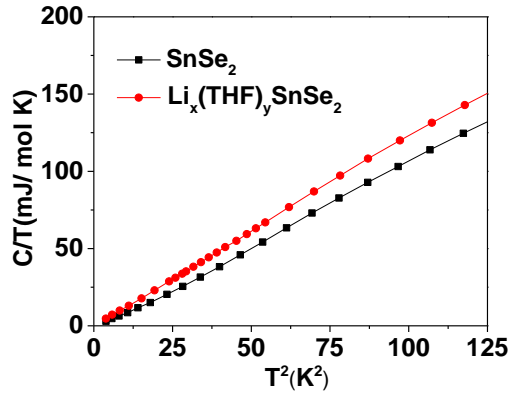

FIG. 6. Heat capacity data of parent and cointercalated SnSe$_2$ samples.

## IV. CONCLUSIONS

In summary, chemical intercalation through both Li doping and the incorporation of organic THF and PC molecules has been carried out in 1T two dimentional SnSe$_2$ phase. We induced superconductivity with values of up to 7.6 K where the transition temperature $T_c$ is found to be significantly enhanced by incorporation of polar organic molecules during the intercalation process. Detailed studies suggests the $T_c$ is almost independent from carrier concentration, but rather correlated with the basal spacing distances. Our results suggest that the charge fluctuation mechanism may play an important role for the unconventional superconductivity present in this system.

## ACKNOWLEDGMENT

This work at University of Texas at Dallas is supported by US Air Force Office of Scientific Research Grant No. FA9550-15-1-0236, FA9550-19-1-0037, and the startup funds from University of Texas at Dallas. The work at AFRL was also supported by an AFOSR grant (LRIR 18RQCOR100) and a grant from the National Research Council. SK and MJK acknowledges the support from Louis Beecherl, Jr. endowed funds.